\newcommand{\bis}{{Bi$_2$S$_3$}}
\newcommand{\sbs}{{Sb$_2$S$_3$}}
\newcommand{\icm}{{cm$^{-1}$}}
\newcommand{\Tr}{{\rm Tr}}

\documentclass[showkeys,showpacs,prb,twocolumn, superscriptaddress,bibnotes]{revtex4}
\usepackage{color}
\usepackage{graphicx}
\usepackage{amsmath}

\makeatletter

\begin{document}

\title{First-principles study of the lattice dynamics of Sb$_2$S$_3$}
\author{Yun Liu}
\affiliation
{Institute of High Performance Computing, Agency
for Science, Technology and Research, 1 Fusionopolis Way, \#16-16
Connexis, Singapore 138632}
\author{Kun Ting Eddie Chua}
\affiliation
{Harvard-Smithsonian Center for Astrophysics, 60 Garden Street, Cambridge, MA 02138, USA}
\author{Tze Chien Sum}
\affiliation
{Division of Physics and Applied Physics, School of Physical \& Mathematical Sciences, Nanyang Technological University, 21 Nanyang Link, Singapore 637371}
\affiliation
{Energy Research Institute at NTU (ERI@N), 1 CleanTech Loop, $\#$06-04, CleanTech One, Singapore 637141}
\affiliation
{Singapore-Berkeley Research Initiative for Sustainable Energy (SinBeRISE), 1 Create Way, Singapore 138602}
\author{Chee Kwan Gan}
\email{ganck@ihpc.a-star.edu.sg}
\affiliation
{Institute of High Performance Computing, Agency
for Science, Technology and Research, 1 Fusionopolis Way, \#16-16
Connexis, Singapore 138632}

\begin{abstract}
We present a lattice dynamics study of orthorhombic antimony sulphide (Sb$_2$S$_3$) obtained using 
density-functional calculations in conjunction with the supercell force-constant method.  
The effect of Born effective charges is taken into account using a mixed-space approach, resulting in the splitting of longitudinal and transverse optical (LO-TO) phonon branches near the zone center. 
Zone-center frequencies agree well with Raman scattering experiments.
Due to the slow decay of the interatomic force constants (IFC), a minimal $2\times4\times2$ supercell ($Pnma$ setting) with 320 atoms is crucial for an accurate determination of the dispersion relations. 
Smaller supercells result in artificial acoustic phonon softening and unphysical lifting of degeneracies along high symmetry directions. 
We propose a scheme to investigate the convergence of the IFC with respect to the supercell sizes. 
The phonon softening can be attributed to the periodic images that affect the accuracy of the force constants, and the truncation of long-ranged forces.
The commensuration of the $\textbf{q}$-vectors with the supercell size is crucial to preserve degeneracies in Sb$_2$S$_3$ crystals.
\end{abstract}
\maketitle

\section{Introduction}
\sbs\ belongs to the group of metal chalcogenides (A$_2$B$_3$, A=As, Sb, Bi and B=S, Se, Te) that form an important class of semiconductors with extensive applications in photovoltaics\cite{C3TC30273C, C2CC17573H} and optoelectronics\cite{Schubert:02}. 
They hold great promise as photovoltaic converters and thermoelectric cooling devices \cite{0022-3727-39-9-014, Mandouh1990} due to their small direct bandgaps, high thermoelectric power, and high absorption coefficient in the visible region \cite{doi:10.1021/jp900302b, doi:10.1021/jz100308q}. 
There is also a surge of interest in using \sbs\ as a solid-state semiconductor-sensitized solar cell to replace the inorganic dye in dye-sensitized solar cells\cite{doi:10.1021/nl101322h, doi:10.1021/jz100308q}. 
\sbs\ has been synthesized to various nanostructured forms, such as nanowires and nanotubes \cite{doi:10.1021/nl2039106, doi:10.1021/nl9002816, B606682H}, which exhibit enhanced ferroelectric, piezoelectric, and conductive properties.

So far, much focus has been placed on the synthesis and electronic properties \cite{BenNasr2011287, Caracas2005, PhysRevB.87.205125, ADFM:ADFM201101103} of \sbs. To complement these known aspects, we present here a study of the lattice vibrational properties of \sbs\ using density-functional theory (DFT). 
Phonon dispersion is one of the fundamental properties of crystals.
The behavior of the branches reflects specific features of the crystal structure and the interactions between the constituent atoms. 
These lattice dynamical properties are indispensable in order to understand the properties of interest for device engineering and design, such as electronic transport and lattice specific heat.

We obtain phonon dispersions of \sbs\ using the supercell force-constant method\cite{0953-8984-9-37-017, 0295-5075-32-9-005, PhysRevLett.74.1791, PhysRevB.73.235214,Gan10v49}, fully taking into account the effect of Born effective charges. 
We note that a similar approach has been employed to study \bis\ in an earlier work\cite{PhysRevB.84.205330}. 
Due to the slow decay of IFC, we also investigate the effects of supercell sizes on the accuracy of the dispersion relations, and propose a scheme to investigate the convergence of the IFC with respect to supercell sizes. 

\section{Methods}

\begin{figure}[h]
\centering
\includegraphics[width=6cm]{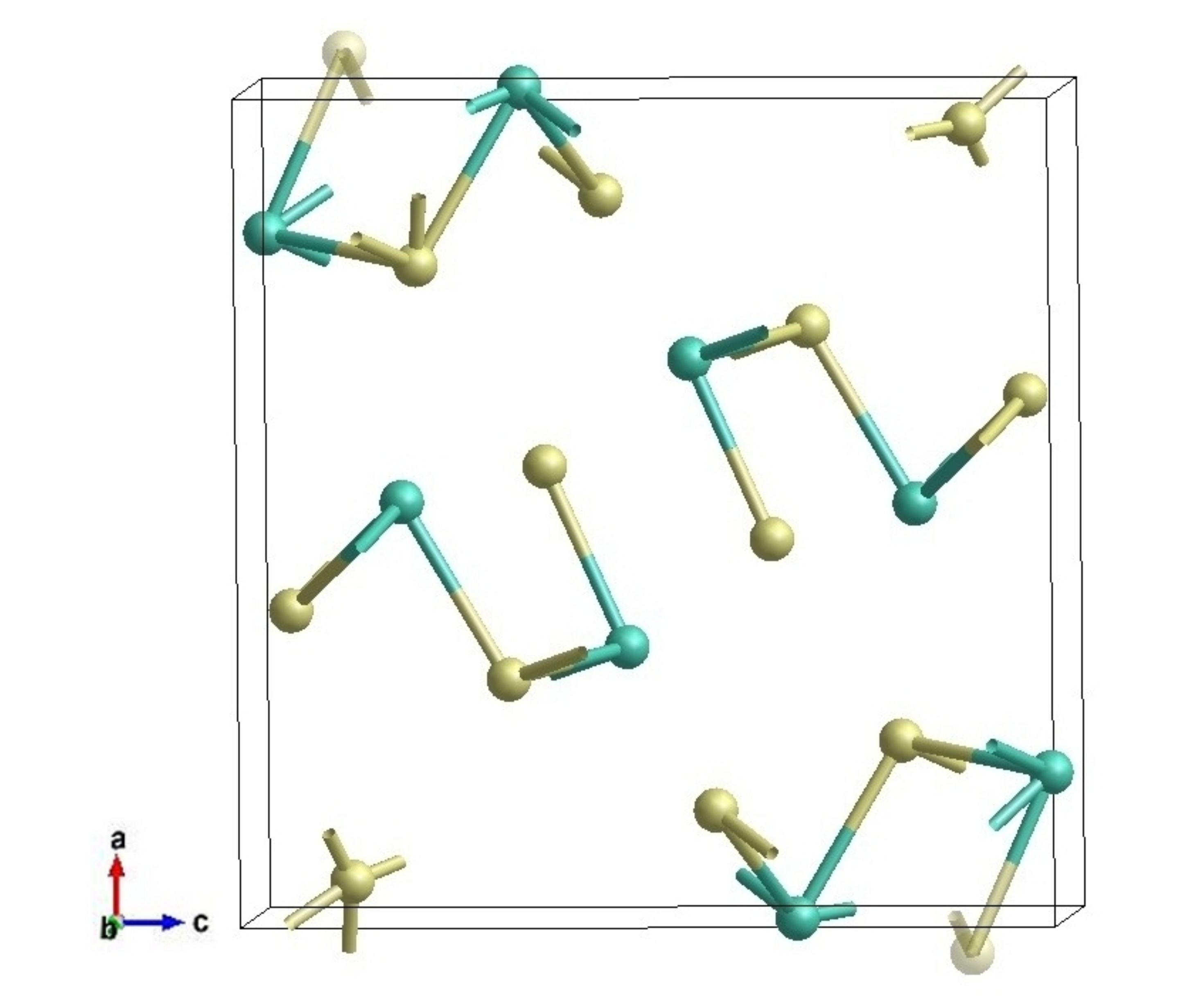}
\caption{Ball-and-stick model of a \sbs\ primitive cell containing 12 S (yellow) and 8 Sb (teal) atoms shown in the $Pnma$ setting. Translucent atoms outside the unit cell have been added to complete the four formula units, each of which contains five inequivalent atoms.}
\label{fig:cell.jpg}
\end{figure}

The orthorhombic phase of \sbs\ belongs to the space group $Pnma$ $ (\#62)$ containing 20 atoms per primitive cell, five of which are inequivalent (Fig. \ref{fig:cell.jpg}). 
DFT calculations are carried out using the {\sc Quantum ESPRESSO}\cite{0953-8984-21-39-395502} suite within the local density approximation (LDA). 
We use pseudopotentials S.pz-bhs.UPF and Sb.pz-bhs.UPF from http://www.quantum-espresso.org.
The electronic wavefunctions are expanded in a plane-wave basis set with kinetic energy cutoff of 75 Ry. 
The Monkhorst-Pack\cite{PhysRevB.13.5188} $k$-point sampling scheme used for Brillouin zone (BZ) integration has divisions of less than $0.03$~\AA$^{-1}$.
The total energies are tested to converge to within 1 meV per atom.

We relax the atomic coordinates and cell dimensions using a Broyden-Fletcher-Goldfarb-Shanno quasi-Newton algorithm to obtain optimized structures with residual forces of less than $10^{-3}$ eV \AA$^{-1}$ and stresses of less than $10^{-4}$ eV \AA$^{-3}$. 
The obtained equilibrium lattice parameters and atomic coordinates of inequivalent atoms are reported in Table \ref{tab:lattconst} which show good agreement with experimental values. 
Note that the theoretical lattice constants $a$, $b$ and $c$ tend to be smaller than the experimental values, which is expected from a LDA calculation\cite{Caracas2005,PhysRevB.87.205125}.

\begin{table*}[ht]
\caption{Equilibrium lattice parameters and inequivalent atomic positions for orthorhombic \sbs\ in the $Pnma$ setting. The experimental data\cite{sb2s3lattice} are included for comparison. Deviations in lattice constants are smaller than 3.8\%.}
\centering
\begin{tabular*}{0.8\textwidth}{@{\extracolsep{\fill}}  c c c | c c c c c c c }
  \hline  \hline
   \multicolumn{3}{c|}{Lattice constants (\AA)} & \multicolumn{7}{c}{Atomic positions}\\ 
  \hline
       & This work & Expt. & & \multicolumn{3}{c}{This work} & \multicolumn{3}{c}{Expt.} \\
      \hline
      &  &  & & $x/a$ & $y/b$ & \multicolumn{1}{c}{$z/c$} & $x/a$ & $y/b$ & $z/c$\\
  $a$ &      11.081 &11.299 & Sb1 (4c) & 0.0200 & 0.25 & 0.6741 & 0.0290 & 0.25 & 0.6738\\                    
  $b$ &      3.831  & 3.831 & Sb2 (4c) & 0.3437 & 0.25 & 0.4677 & 0.3503 & 0.25 & 0.4641\\        
  $c$ &  10.805 &11.227  & S1 (4c) & 0.0508 & 0.25 & 0.1282 & 0.0493 & 0.25 & 0.1226\\  
      $b/a$ & 0.346  &0.339  & S2 (4c) & 0.3738 & 0.25 & 0.0567 & 0.3745 & 0.25 & 0.0612\\ 
      $c/a$ & 0.975  &0.994  & S3 (4c) & 0.2133 & 0.25 & 0.8068 & 0.2077 & 0.25 & 0.8071\\
 \hline \hline
\end{tabular*}
\label{tab:lattconst}
\end{table*}

To obtain the full phonon dispersion within the BZ, we adopt a direct, supercell force-constant approach\cite{0295-5075-32-9-005, PhysRevLett.74.1791, PhysRevB.73.235214,Gan10v49, RevModPhys.74.11}. 
We note that density-functional perturbation theory (DFPT)\cite{Baroni01v73,Gonze97v55} may also be employed to obtain the phonon dispersions. 
The relative merits of these two methods have been discussed in  Ref.[\citenum{Baroni01v73}]. 
In the supercell force-constant approach, the $s$th atom in the primitive cell is displaced from its equilibrium position in the $\alpha$ directions by $\pm\delta^{\alpha}_{s}=\pm0.015$ \AA. 
The forces acting on the  $u$th atom in the supercell, $F^{\beta}_{u}(\pm\delta^{\alpha}_{s})$ are calculated using the Hellmann-Feynman theorem. $\alpha$ and $\beta$ denote the three spatial directions. 
We then use a finite central-difference scheme to evaluate the matrix elements of the IFC $\phi$ as
\begin{equation}
 \phi^{\alpha\beta}_{su} = - \left[ \frac{F^{\beta}_{u}(+\delta^{\alpha}_{s})-F^{\beta}_{u}(-\delta^{\alpha}_{s})}{2\delta^{\alpha}_{s}} \right].
\end{equation}
The dynamical matrix at a $\textbf{q}$-point is obtained by summing the contributions from all atoms in the supercell
\begin{equation}
D^{\alpha\beta}_{st}(\textbf{q}) = \frac{1}{\sqrt{M_s M_t}} \sum_{\mathbf{R}} \phi^{\alpha\beta}_{st}(\mathbf{R})\ e^{i\mathbf{q}\cdot\mathbf{R}}
\end{equation}
where $s$ and $t$ run over all atoms in the primitive cell, $M_s$ is the mass of the $s$th atom and \textbf{R} is a lattice translation vector. 
$\phi^{\alpha\beta}_{st}(\mathbf{R})$ denotes the IFC between the $s$th atom in the primitive cell, and another atom in the supercell that is the image of the $t$th atom in the primitive cell under \textbf{R}. 
The diagonalization of $D$(\textbf{q}) then yields the phonon frequencies at $\textbf{q}$.

\subsection{Symmetry Reduction}

In order to reduce the number of static DFT calculations, we use the symmetry properties of the crystals to transform the forces\cite{0295-5075-32-9-005}. 
Only the inequivalent atoms within the primitive cell are displaced to find the forces $F_{su}$ between the $s$th atom in the primitive cell and $u$th atom in the supercell, which can be represented by a 3 $\times$ 3 matrix. 
To obtain the forces between an equivalent atom and all other atoms in the supercell, we use the space group operation $S$ that maps the $s$th inequivalent atom to the $p_s$th equivalent atom in the primitive cell. 
The forces between $p_s$th and $p_u$th atoms can be simply calculated as
\begin{equation}
  F_{p_s p_u} = \textbf{G}(S)\ F_{su}\ \textbf{G}(S^{-1})
\label{equation:pointgroup}
\end{equation}
where $\textbf{G}(S)$ represents the point group part of $S$ in Cartesian coordinates. This approach allows us to displace five atoms in the three spatial directions rather than all 20 atoms in the primitive cell, resulting in substantial saving of the total calculation time.

\subsection{Non-analytical correction for $\textbf{q}\rightarrow0$}

Due to the polar character of \sbs, the long-range dipole-dipole interaction gives rise to a macroscopic electric field that affects longitudinal optical (LO) phonon modes and not the transverse optical (TO) modes\cite{Zhong94v72}. The LO-TO splitting depends on the direction from which one approaches the $\Gamma$ point in the BZ. 
This effect is reflected in the non-vanishing Born effective charge tensor \textbf{Z$^*$}, taking the form of a non-analytical contribution $\tilde{D}^{\alpha\beta}_{st}$ to the dynamical matrix\cite{Cochran1962447,PhysRevB.43.7231, Baroni01v73,Gonze97v55} in the limit \textbf{q}$\rightarrow0$:
\begin{eqnarray}
  \tilde{D}^{\alpha\beta}_{st}(\mathbf{q}\rightarrow 0) &=& 
  \frac{4 \pi e^2}{\Omega \sqrt{M_s M_t}} \frac{\sum_{\gamma}
   Z^{*\gamma\alpha}_{s} q_{\gamma} \sum_{\nu} 
   Z^{*\nu\beta}_{t} q_{\nu}}
   {\sum_{\gamma,\nu}q_{\gamma} \epsilon^{\gamma 
   \nu}_{\infty}q_{\nu}} \nonumber\\  
  &=&  \frac{4 \pi e^2}   {\Omega \sqrt{M_s M_t}} \frac{(\mathbf{q} \cdot \mathbf{Z}{^*_s})_\alpha   (\mathbf{q} \cdot \mathbf{Z}{^*_t})_\beta} {\mathbf{q} \cdot \boldsymbol{\epsilon}^\infty  \cdot \mathbf{q}} 
\label{equation:nonanalytical}
\end{eqnarray}
where $\Omega$ is the volume of the primitive cell, $e$ is the elementary charge, $\boldsymbol{\epsilon}^\infty$ is the high-frequency dielectric tensor and \textbf{Z}$^*_s$ is the Born effective charge tensor for the $s$th atom. \textbf{Z}$^*$ and $\boldsymbol{\epsilon}^\infty$ 
may be calculated using density-functional perturbation theory (DFPT) \cite{Baroni01v73,Gonze97v55}. 

In order to include the non-analytical correction in the phonon dispersion, we add\cite{Wang10v22} a correction factor $\tilde{\varphi}$ to the real space force-constant $\phi$:
\begin{equation}
 \Phi^{\alpha\beta}_{st}(\mathbf{R}) = \phi^{\alpha\beta}_{st}(\mathbf{R}) + \tilde{\varphi}^{\alpha\beta}_{st}
 \label{equation:Phi}
\end{equation}
where $\Phi(\mathbf{R})$ is the corrected real-space inter-atomic force-constant matrix. We may calculate the correction factor $\tilde{\varphi}^{\alpha\beta}_{st}$ by imposing the condition that 
\begin{equation}
\lim_{\mathbf{q} \to 0} \frac{1}{\sqrt{M_s M_t}}  \sum_{\mathbf{R}}  \tilde{\varphi}^{\alpha\beta}_{st} e^{i\mathbf{q}\cdot\mathbf{R}} =  \tilde{D}^{\alpha\beta}_{st}(\mathbf{q}\rightarrow 0),
\end{equation}
from which we obtain 

\begin{eqnarray}
     \tilde{\varphi}^{\alpha\beta}_{st} 
     = \frac{1}{N} \frac{4 \pi e^2} {\Omega } \frac{(\mathbf{q} \cdot \mathbf{Z}{^*_s})_\alpha   (\mathbf{q} \cdot \mathbf{Z}{^*_t})_\beta} {\mathbf{q} \cdot \boldsymbol{\epsilon}^\infty  \cdot \mathbf{q}} 
   \label{equation:varphi}
\end{eqnarray}
where $N$ is the number of primitive cells in the supercell. The corrected force-constant matrix $\Phi$ (Eqn.~\ref{equation:Phi}) is finally used to calculate the phonon frequencies. 

\section{Results and Discussion}

\subsection{Born effective charges}

\begin{table}[h]
\caption{Non-zero components of the Born effective charge tensor $\textbf{Z}^*$ and high-frequency dielectric tensor $\boldsymbol{\epsilon}^\infty$ of \sbs. 
Components of the five inequivalent atoms are shown for $\textbf{Z}^*$.
The values of the equivalent atoms are the same up to a sign and can be calculated using the transformation relation according to Eqn. \ref{equation:pointgroup}.}
\centering
\begin{tabular*}{8.25cm}{@{\extracolsep{\fill}} c c c c c c }  
  \hline \hline
  & xx & yy & zz & xz & zx \\
  \hline
   $\epsilon^\infty$ & 9.33 & 18.7 & 13.0 & 0 & 0 \\ 
   Z$^*$(Sb1) & 2.89  & 5.62  & 7.36  & 0.07 & 1.53  \\
   Z$^*$(Sb2) & 3.33  & 7.25  & 4.50  & 0.28 & 0.09  \\
   Z$^*$(S1)  & -2.35 & -4.18 & -4.07 & 1.07 & 0.83  \\
   Z$^*$(S2)  & -1.83 & -4.80 & -4.44 & -0.45 & -0.33  \\
   Z$^*$(S3)  & -2.03 & -3.90 & -3.36 & -0.20 & -1.34  \\
  \hline \hline
\end{tabular*}
\label{tab:BEC}
\end{table}

The Born effective charge Z$^{*\alpha\beta}_s$ is the first derivative of the macroscopic polarization along the $\alpha$ direction with respect to the displacement of $s$th atom along the $\beta$ direction. 
This quantity is calculated using linear response theory at the zone center ($\Gamma$), and the non-zero values are shown in Table \ref{tab:BEC}. 
The diagonal elements are different for each Sb and S atom, and off-diagonal elements are present, showing considerable anisotropy in the system.

The formal valence charges for Sb and S are $+3$ and $-2$. Our calculation shows maximum effective charges of $+7.36$ and $-4.80$ for Sb and S respectively. 
From the study of ferroelectric compounds, it has been suggested that ions with effective charges close to the formal valence charge behave as closed-shell ions. 
Conversely, the presence of covalent character in the bonds causes a large amount of delocalized charge to flow through the structure during lattice displacements{\cite{Zhong94v72,Posternak94v12,Amritendu10v22}. 
In \sbs, the significantly larger effective charges suggest that there is substantial covalent character in the bonds. 
Sb atoms are able to donate electrons to S atoms during lattice displacements, and hence increase the magnitude of their respective Born effective charges. 
These results agree with X-ray photoelectron spectroscopy studies that describe the bonding in \sbs\ as tight covalent\cite{doi:10.1080/01411590290020448}.

We also obtain the high-frequency dielectric tensor $\boldsymbol{\epsilon}^\infty$ which is diagonal as shown in Table \ref{tab:BEC}. The anisotropy of $\boldsymbol{\epsilon}^\infty$ is an indication of the anisotropy of \sbs\ structure.

\subsection{Zone-center phonons}

\begin{table}[ht]
\caption{Zone-center phonon modes in \sbs\ obtained using the supercell 
force-constant method, DFPT and Raman scattering spectroscopy. Only phonon modes with Raman data are shown. A complete list of Raman active modes are listed in the Supplementary Information.}
\centering
\begin{tabular*}{0.41\textwidth} {@{\extracolsep{\fill}}c r r r r}
  \hline \hline  
  Raman & Intensity & DFPT & Supercell & Expt.\cite{PhysRevB.84.205330} \\
  mode & \AA$^4$ u$^{-1}$ & (\icm) & (\icm)  & (\icm)  \\
  \hline
  B$_{1g}$ & 267 & 47.7 & 47.7& 43\\
  B$_{3g}$ & 1320 & 50.8& 50.8 & 52\\
  A$_{g}$ & 475 & 54.3 & 54.2 & 51\\
  B$_{3g}$ & 1140 & 69.1& 69.0& 60\\
  A$_{g}$ & 1160 & 74.5& 74.2& 72\\
   B$_{2g}$ & 573 & 99.1& 99.1& 91\\
   A$_{g}$ & 187 & 100.0& 100.0 & 101\\
   B$_{2g}$ & 9.3& 124.1& 125.0& 128\\
   A$_{g}$ & 9970& 196.7& 197.4& 192\\
   B$_{1g}$& 1590 & 208.2& 208.2 & 207\\
   B$_{3g}$ & 1390 & 231.4& 231.4 & 239\\
   A$_{g}$& 4500& 251.0 &251.0 &256\\
   A$_{g}$ & 19100 & 277.9& 278.0& 283\\
  \hline\hline
\end{tabular*}
\label{tab:frequencies}
\end{table}

Since $Pnma$ is a centrosymmetric space group, the Raman and infra-red (IR) modes of \sbs\ are mutually exclusive, i.e., a mode cannot be simultaneously Raman and IR active. There are 60 phonon modes at $\Gamma$ that respect the $D_{2h}$ point group symmetry:
\begin{equation}
  \Gamma = 3\ \Gamma_{\text{acoustic}} + 30\ \Gamma_{\text{Raman}} + 22\ \Gamma_{\text{IR}} + 5\ \Gamma_{\text{silent}} \nonumber\\
\end{equation}
3 are acoustic phonon modes ($\Gamma_{\text{acoustic}} = B_{1u} +  B_{2u} + B_{3u}$). Of the optical phonon modes, 30 are Raman active ($\Gamma_{\text{Raman}} = 10 A_g + 5 B_{1g} + 10 B_{2g} + 5 B_{3g}$), 22 are IR active ($\Gamma_{\text{IR}} = 4 B_{1u} + 9 B_{2u} + 9 B_{3u}$), and 5 are optically silent ($\Gamma_{\text{silent}} = 5 A_u$). 
The Sb and S atoms have the site symmetry $C_{s}$ that restrict their motions within the $xz$ plane for the $A_g$, $B_{2g}$, $B_{1u}$ and $B_{3u}$ modes, and along the $y$ axis for the $B_{1g}$, $B_{3g}$, $A_u$ and $B_{2u}$ modes.
This symmetry is also consistent with the anisotropy reflected in the Born effective charges, where the only non-zero off-diagonal terms are due to the coupling in the $x$ and $z$ axis. 
The restriction means that motion in the $xz$ plane is independent of the motion along $y$ direction, and hence the displacements of atoms along $y$ direction will not result in any polarization along $x$ or $z$ directions, and vice versa.

Currently there is no systematic study to assign the experimentally observed Raman modes in \sbs. Sereni {\it et al.}\cite{sereni:1131} recently performed a polarization-dependent Raman scattering study on single crystal samples of \sbs\ in the 90$^\circ$ and 180$^\circ$ geometries. 
Contributions from $B_{1g}$ spectrum were found in the experimental data for the $A_g$ measurements. They also reported first-principles calculations of the zone-center phonons, but theoretical Raman scattering intensities were lacking. 
This makes the discrimination between $A_g$ and $B_{2g}$ modes, and that between $B_{1g}$ and $B_{3g}$ modes very challenging, as the effect of microtwinning makes the $a$ and $c$ directions indistinguishable \cite{PhysRevB.57.2872}.
We present here the non-resonant Raman scattering coefficients which are computed from the second order derivative of the electronic density matrix with respect to a uniform electric field as implemented in {\sc Quantum ESPRESSO}\cite{PhysRevLett.90.036401}. 
To assign the modes, the phonon frequencies, scattering coefficients and the space group symmetries are taken into consideration. 
We find that 13 phonon modes are in good agreement with the experimental values, as shown in 
Table~\ref{tab:frequencies}.
However, experimental studies on additional scattering geometries are needed to provide a comprehensive assignment. 

\begin{figure*}[ht]
\centering
\includegraphics[width=\textwidth,clip]{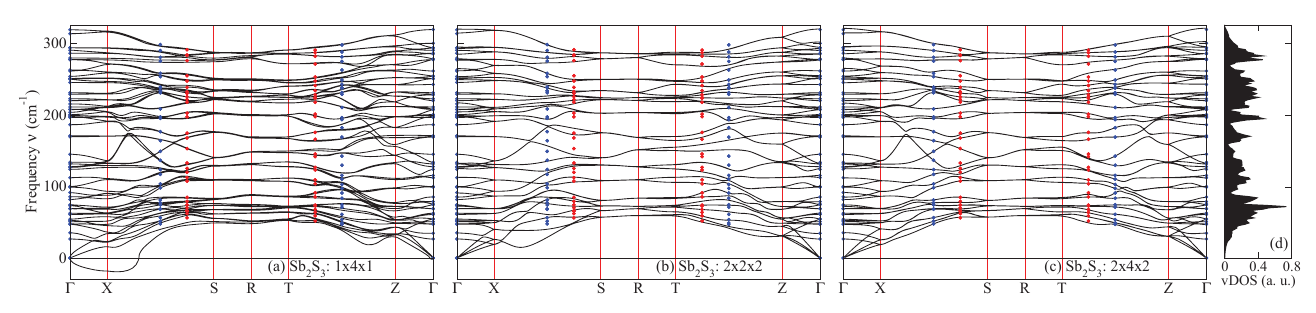}
\caption{First-principles phonon dispersions along high symmetry directions
for orthorhombic \sbs\ calculated using the supercell force-constant
method for $1\times4\times1$, $2\times2\times2$ and $2\times4\times2$
supercells.  The selected $\textbf{q}$-points are $\Gamma=(0,0,0)$,
$X=(\frac{1}{2},0,0)$, $S=(\frac{1}{2},\frac{1}{2},0)$,
$R=(\frac{1}{2},\frac{1}{2},\frac{1}{2})$,
$T=(0,\frac{1}{2},\frac{1}{2})$, and $Z=(0,0,\frac{1}{2})$.  Blue circles
are phonon frequencies calculated from DFPT at $\textbf{q}$-points
commensurate with $2\times4\times2$ supercell, while red circles are
frequencies at $\textbf{q}$-points non-commensurate with any of the
supercells.  Imaginary frequencies (represented by negative frequencies)
are present in $1\times4\times1$ supercell. Degeneracies are preserved
in the $2\times2\times2$ and $2\times4\times2$ supercells but lost in the
$1\times4\times1$ supercell.
}
 \label{fig:dispersions}
\end{figure*}

\subsection{Phonon dispersions}

Periodic images of displaced atoms can exert sizable effects during
the calculation of forces, hence decreasing the accuracy of the IFC.
This is not an issue for $\textbf{q}$-vectors that are commensurate
with the supercell, since the phonon frequencies calculated at these
points are exact using the supercell method.  In order to reduce the
effects of periodic images on non-commensurate $\textbf{q}$-points, a
huge supercell has to be used.  However, large primitive cells such as
that of \sbs\ place a computational constraint on the largest supercells
that can be used in DFT calculations.  As a result, we investigate the
effect of supercell sizes on the phonon dispersions by using sizes of
$1\times 4\times 1$ containing 80 atoms , $2\times 2\times 2$ containing
160 atoms and $2\times 4\times 2$ containing 320 atoms. 
We plot the dispersion relations
along the high symmetry directions $\Gamma
\rightarrow X \rightarrow S \rightarrow R \rightarrow T \rightarrow
Z \rightarrow \Gamma$ as shown in Fig.~\ref{fig:dispersions}, where we have used the ${\bf q}$-vector convention
in Refs.\cite{Setyawan2010299,Rao1982Pramana19}. Figs.~\ref{fig:dispersions}(c) shows our most
accurate results from the $2\times4\times2$ supercell calculation.

There are in general 60 phonon modes present in the dispersions,
as evident in the $X \rightarrow \Gamma$ and $Z \rightarrow \Gamma$
directions.  Double degeneracy occurs along the high symmetry lines
$X\rightarrow S$ and $R\rightarrow T\rightarrow Z$, resulting in only
30 distinct phonon frequencies. Along $S\rightarrow R$, the phonon
frequencies are quadruply degenerate and only 15 frequencies are
present.  These degeneracies are preserved in the $2\times2\times2$
and $2\times4\times2$ results whereby the zone boundary points ($X$,
$S$, $R$, $T$ and $Z$) are commensurate with the supercell sizes, i.e.,
\begin{equation}
   \textbf{q}\cdot\textbf{L}_i=2\pi n_i
\label{equation:exact} \end{equation} where $n_i$ is an integer
and $\textbf{L}_i$ are the three supercell lattice vectors.
It is interesting
to note that commensuration at these points helps to ensure that the
entire dispersion along the high symmetry directions also have the
correct degeneracies.  In contrast, degeneracies are lifted in smaller
supercell sizes (Figs. \ref{fig:dispersions}(a)), as the zone boundary
points are not commensurate with the supercell.  Commensuration is thus
extremely important in low symmetry crystals such as \sbs\ to preserve
the correct degeneracies as opposed to high symmetry crystals which
may retain degeneracies even when the commensuration criteria are not
met. The conclusion is not specific to \sbs\ as we also observed the
same behavior for \bis\ as shown in the Supplementary Information.

\begin{figure*}[ht]
\centering
\includegraphics[width=13cm]{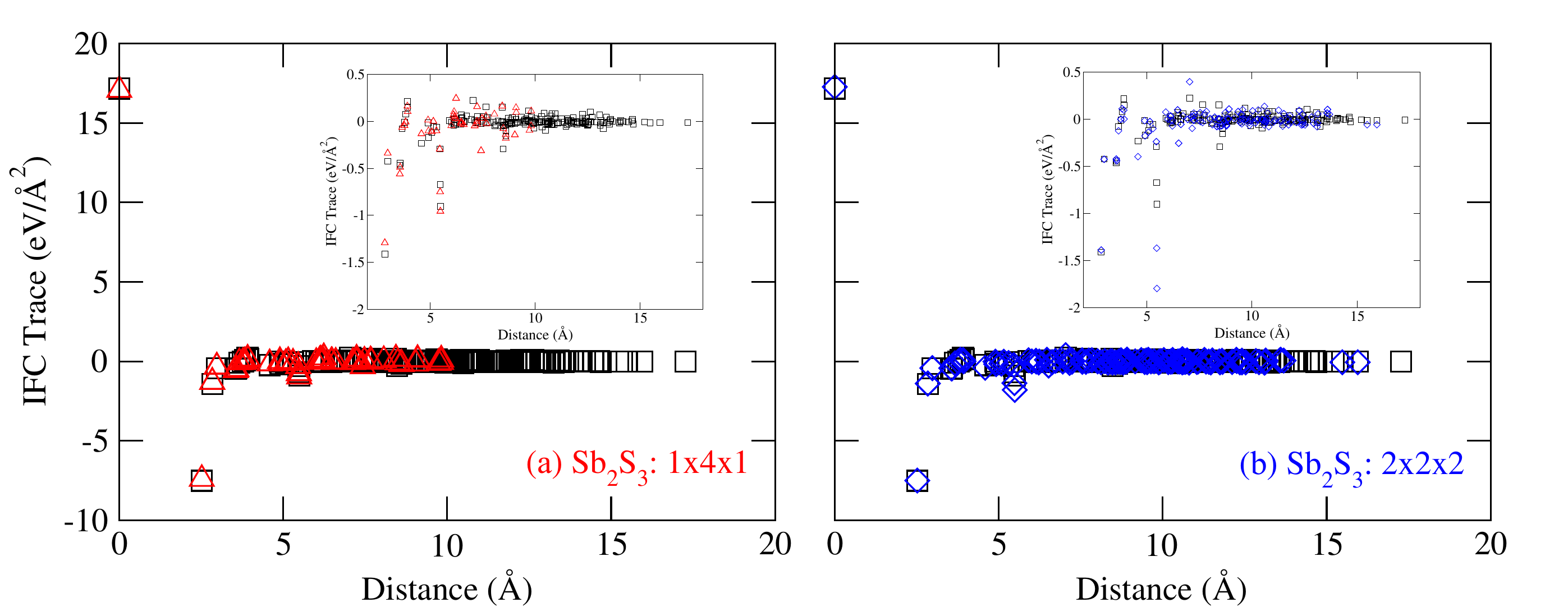}
\caption{$\Tr(\Phi_{su})$ as a function of $r_{su} = |\textbf{r}_s - \textbf{r}_u|$, the distance between the $s$th and $u$th atom. 
The $s$th atom is the inequivalent S1 sulphur atom, and $u$ runs through all the atoms within the supercell. 
The red triangles are data from $1\times4\times1$ supercell, blue diamonds from $2\times2\times2$, and black squares for $2\times4\times2$. The insets show significant interactions beyond the on-site and nearest neighbor atoms.}
\label{fig:fconstant}
\end{figure*}

As shown in Fig.~\ref{fig:dispersions}, the zone-center phonon frequencies
of \sbs\ differ in the $X\rightarrow \Gamma$ and $Z\rightarrow
\Gamma$ directions due to macroscopic electric fields in the polar
crystals (the so-called LO-TO splitting). 
$D=(\frac{1}{2},\frac{1}{4},0)$
and $B=(0,\frac{1}{4},\frac{1}{2})$ are only commensurate with
$2\times4\times2$ supercell, and therefore the agreement between DFPT and
supercell results are excellent. There are, however, huge discrepancies
in the smaller supercells, showing that the dispersions are not accurate.
$(\frac{1}{2},\frac{3}{8},0)$ and $(0,\frac{3}{8},\frac{1}{2})$ are
not commensurate with any of the supercell sizes, but the results
from the supercell force-constant method and DFPT show a good agreement in
Figs. \ref{fig:dispersions}(c).

Apart from affecting the degeneracies, small supercell sizes also
result in the artificial softening of phonon branches, as can be seen
in Figs.~\ref{fig:dispersions}(a) and (b). 
The softening becomes so
severe in the $1\times4\times1$ supercell that imaginary frequencies
are introduced in the acoustic phonon modes.
Soft modes are also displayed in the dispersions of $1\times 1\times 1$ and $1\times 2\times 1$ supercells
(see the Supplementary Information).  

One of the biggest challenges in phonon calculations 
is to determine whether the soft modes are artificial
(an artefact of the numerical methods, etc) or 
genuine that are associated with unstable lattice structures and phase transitions\cite{Parlinski97v78,Duan11v84}.
The occurrence of artificial phonon soft modes found in this work should
serve as a caution that convergence with respect to supercell size must be
checked carefully when using the supercell force-constant method.
We conclude that softening is totally absent in the
dispersions obtained with a $2\times4\times2$ supercell for \sbs.

The vibrational density of states (vDOS) using the $2\times4\times2$
supercell result is shown in Fig. \ref{fig:dispersions}(d). A dense
$10\times30\times10$ $k$-point mesh is used to sample the BZ and the
effect of LO-TO splitting is included.

\subsection{IFC analysis}

We now propose a scheme to analyze the IFC in order to understand the origin of the phonon softening. 
For each pair of atoms, a $3\times3$ force constant matrix is obtained, with elements corresponding to movements of the each atom along the Cartesian directions. 
As a measure of the strength of this interaction, we use the trace of the IFC tensor, Tr($\Phi_{su}$) which has the advantage of being independent of the coordinate system used. 
The decrease of Tr($\Phi_{su}$) with increasing interatomic distance $r_{su}$ suggests a suitable range for the interatomic forces\cite{0370-1301-70-12-305}.

Fig. \ref{fig:fconstant} shows the decay in Tr($\Phi_{su}$) as a function of distances for different supercell sizes. The $s$th atom is the inequivalent S1 sulphur atom (Table \ref{tab:lattconst}), and $u$ runs through all atoms within the supercell. Although not shown here, similar features are observed when the $s$th atom is replaced by other inequivalent atoms. 
Although the interactions are dominated by the on-site and the nearest neighbor terms, there are significant non-zero contributions at large distances which show the long ranged interactions in the crystals.
The lattice dynamics of \sbs\ thus cannot be approximated by a simple linear chain model considering only a few nearest neighbor interactions. To correctly describe the dispersions, a large supercell is needed to reduce the effect of periodic images and to capture the interactions at large distances.

Due to the small sizes of $1\times4\times1$ and $2\times2\times2$ supercells, many values of IFC differ from those of $2\times4\times2$ by a few orders of magnitude, as can be seen in Fig. \ref{fig:logtr}. 
In addition, many interactions beyond $\sim 6$ \AA\ are not captured in the smaller supercells. 
We believe that these effects destabilize the crystal structure and result the softening of the acoustic phonons. The softening is more pronounced in $1\times4\times1$ supercells and less so in $2\times2\times2$, due to the larger size of the supercell that reduces the effect of the periodic images and captures more long ranged interactions. 

\begin{figure}[ht]
\centering
\includegraphics[width=8cm]{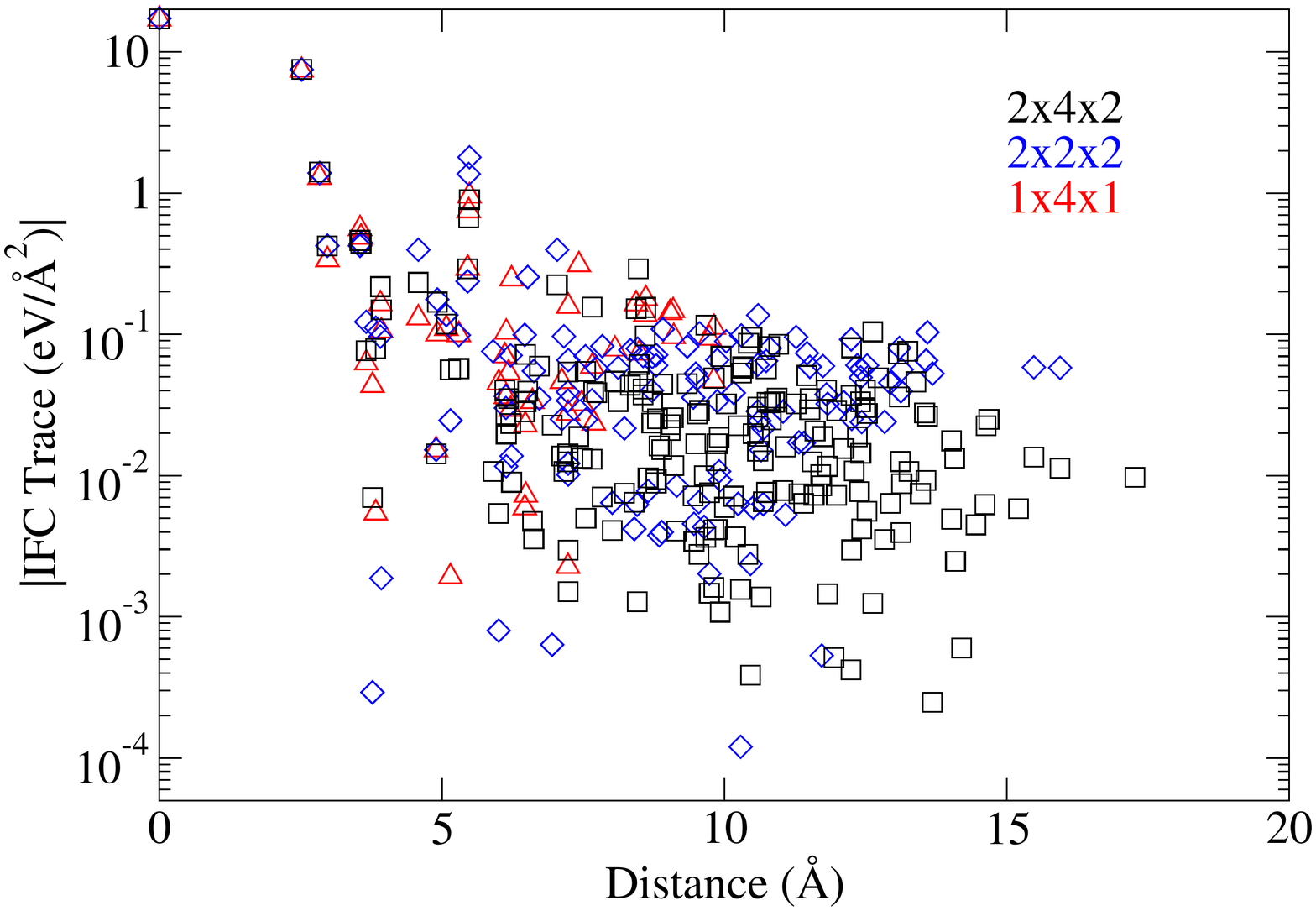}
\caption{The logarithm of the absolute value of Tr($\Phi_{su}$) as a function of the distance $r_{su}$, for the various supercell sizes of \sbs. The $s$th atom is the inequivalent S1 sulphur atom, and $u$ runs over all atoms within the supercell. The values of the IFC differ by up to a few orders of magnitude between the smaller supercells and $2\times4\times2$ supercell.}
\label{fig:logtr}
\end{figure}

\section{Conclusion}

To the best of our knowledge, the phonon dispersion of \sbs\ is obtained
for the first time through a systematic lattice dynamics study on low
symmetry crystals using the supercell force-constant method.  The Born
effective charges give rise to LO-TO splitting at the zone center and
elucidate the covalent character of the bonds.  Both the high frequency
dielectric tensor and Born effective charges show considerable anisotropy
of the crystals.  
The use of small supercell sizes
results in the softening of the phonon modes that is inconsistent with
experiments.  We attribute this to the effect of the periodic images on
the force constants, as well as the truncation of long ranged interactions.
We found that a
minimal $2\times 4 \times 2$ supercell ($Pnma$ setting) is required for an
accurate determination of the dispersion relations of \sbs.  Our results
suggest that when using the supercell force-constant method, the supercell
size has to be tested with other parameters such as the kinetic energy
cut-off, the Brillouin-zone sampling or the self-consistent convergence
criteria especially when dealing with low symmetry systems such as \sbs.

\section*{Acknowledgements}
The authors thank Peter Haynes of Imperial College London for useful discussions and pointing out Ref.[\citenum{Caracas2005}]. Y.L and K.T.E.C acknowledge the financial support from the NSS programme, Singapore. T.C.S acknowledges the support by the following research grants: NTU start-up grant (M4080514); SPMS collaborative Research Award (M4080536); and the Singapore-Berkeley Research Initiative for Sustainable Energy (SinBeRISE) CREATE Programme. The authors gratefully acknowledge the use of resources at the A$\ast$STAR Computational Resource Centre, Singapore.

\footnotesize{
\providecommand*{\mcitethebibliography}{\thebibliography}
\csname @ifundefined\endcsname{endmcitethebibliography}
{\let\endmcitethebibliography\endthebibliography}{}

}

\end{document}